\documentclass[traditabstract]{aa}
\usepackage{natbib}
\usepackage{epsfig}

\newcommand{\ms}{\mbox{m\,s$^{-1}$}}
\newcommand{\teff}{\mbox{$T_{\rm eff}$}}
\newcommand{\logg}{\mbox{$\log g$ }}

\begin{document}

\title{Detection of a transit of the super-Earth 55\,Cnc\,e with warm~{\it Spitzer}\thanks{The photometric time series used in this work are available in electronic form at the CDS via anonymous ftp to cdsarc.u-strasbg.fr (130.79.128.5) or via http://cdsweb.u-strasbg.fr/cgi-bin/qcat?J/A+A/}}
\author{B.-O. Demory$^{1}$, M. Gillon$^{2}$, D. Deming$^3$, D. Valencia$^1$, S. Seager$^1$, B. Benneke$^1$, C. Lovis$^4$, P. Cubillos$^5$, J. Harrington$^5$, K. B. Stevenson$^5$, M. Mayor$^4$, F. Pepe$^4$, D. Queloz$^4$, D. S\'egransan$^4$, S. Udry$^4$} 

\offprints{demory@mit.edu}
\institute{
$^1$ Department of Earth, Atmospheric and Planetary Sciences, Department of Physics, Massachusetts Institute of Technology, 77 Massachusetts Ave., Cambridge, MA 02139, USA\\
$^2$ Institut d'Astrophysique et de G\'eophysique,  Universit\'e de Li\`ege,  All\'ee du 6 Ao\^ut 17,  Bat.  B5C, 4000 Li\`ege, Belgium \\
$^3$ Department of Astronomy, University of Maryland, College Park, MD 20742-2421, USA\\
$^4$ Observatoire de Gen\`eve, Universit\'e de Gen\`eve, 51 Chemin des Maillettes, 1290 Sauverny, Switzerland\\
$^5$ Planetary Science Group, Department of Physics, University of Central Florida, Orlando, FL 32816-2385, USA\\
}

\date{Received 3 May 2011 / Accepted 31 July 2011}
\authorrunning{B.-O. Demory et al.}
\titlerunning{Detection of a transit of 55\,Cnc\,e with warm {\it Spitzer}}
\abstract{
We report on the detection of a transit of the super-Earth 55\,Cnc\,e 
with warm {\it Spitzer} in IRAC's 4.5$\mu$m band. Our MCMC analysis includes an extensive 
modeling of the systematic effects affecting warm {\it Spitzer} photometry, and yields a transit 
depth of $410 \pm 63$ ppm, which translates to a planetary radius of $2.08^{+0.16}_{-0.17}\: R_\oplus$ 
as measured in IRAC 4.5$\mu$m channel. A planetary mass of $7.81_{-0.53}^{+0.58}\: M_\oplus$ is derived 
from an extensive set of radial-velocity data, yielding a mean planetary density of $4.78_{-1.20}^{+1.31}$ g\,cm$^{-3}$. 
Thanks to the brightness of its host star ($V=6$, $K=4$), 55\,Cnc\,e is 
a unique target for the thorough characterization of a super-Earth orbiting around a solar-type star.
   }

\maketitle

\section{Introduction}
Radial velocity (RV), microlensing and transit surveys have revealed the existence in our Galaxy of a large population of planets with a mass of a few to $\sim$20 Earth masses (Lovis et al. 2009; Sumi et al. 2010; Borucki et al. 2011). Based on their mass (or minimum mass for RV planets), these planets are loosely classified  as ``super-Earths" ($M_p \le10\: M_\oplus$) and ``Neptunes" ($M_p > 10\: M_\oplus$). This classification is  based on the theoretical limit for gravitational capture of H/He, $\sim$10 $M_\oplus$ (e.g., Rafikov 2006), and thus implicitly assumes that  Neptunes are predominantly ice giants with a significant H/He envelope, and that most super-Earths are massive  terrestrial planets. Still, the diversity of this planetary population is probably much larger than sketched by this simple division, as we can expect from the stochastic nature of planetary formation. 
 
The first transit of one of these low-mass planets, GJ\,436\,b, was detected in 2007 (Gillon et al. 2007).
Thanks to its transiting nature, the actual mass ($M_p= 23.2 \pm 0.8\: M_\oplus$) and radius ($R_p=4.22 \pm 0.10\: R_\oplus$) of GJ\,436\,b could be accurately determined (Torres, 2007), indicating for this ``hot Neptune" a mass, radius and density indeed very similar to the ice giant planets Uranus and Neptune. More recently, several other transiting low-mass planets were detected. While many more planet candidates detected by the $Kepler$ mission are waiting for confirmation (Borucki et al. 2011), the first confirmed low-mass transiting planets already show a large diversity. Some of these planets, like HAT-P-11\,b (Bakos et al. 2010) and Kepler-4\,b (Borucki et al. 2010b), are similar to Neptune and GJ\,436\,b.  Kepler-11\,c (Lissauer et al. 2011) seems to be a smaller version of Neptune, while HAT-P-26\,b (Hartman et al. 2010) has a much lower density (0.4 $\pm$ 0.10 g\,cm$^{-3}$ $vs$ 1.64 g\,cm$^{-3}$ for Neptune) that is consistent with a significantly larger H/He fraction. The super-Earths CoRoT-7\,b (L\'eger et al. 2009, Hatzes et al. 2010) and Kepler-10\,b (Batalha et al.  2011) are probably massive rocky planets formed in the inner part of their protoplanetary disks. The super-Earth GJ\,1214\,b (Charbonneau et al. 2009) is still mysterious in nature. Its large radius ($R_p = 2.44 \pm 0.21\: R_{\oplus}$, Carter et al. 2011) suggests a  significant gaseous envelope that could  originate from the  outgassing of the rocky/icy surface material of a terrestrial planet or that could be of primordial origin, making it a kind of ``mini-Neptune" (Rogers \& Seager 2010). Recent transit transmission spectrophotometric measurements for GJ\,1214\,b seem to rule out a cloud-free atmosphere composed primarily of hydrogen (Bean et al. 2010, D\'esert et al. 2011), but more atmospheric measurements are needed to determine the exact nature of its envelope. The case of GJ\,1214\,b shows nicely that understanding the true nature of a low-mass exoplanet could require not only precise measurements of its mass and radius, but also a study of its atmospheric properties. 

Among all the low-mass transiting exoplanets detected so far, only GJ\,1214\,b and GJ\,436\,b, and to a lesser extent HAT-P-11\,b and HAT-P-26\,b, orbit around stars small enough and bright enough in the infrared to make possible a thorough atmospheric characterization with existing or future facilities like JWST (e.g., Shabram et al. 2011). Improving our understanding of the low-mass planet population orbiting around solar-type stars requires that such planets are detected in transit in front of  much nearer/brighter host stars than the targets of surveys like CoRoT (Barge et al. 2008) or {\it Kepler} (Borucki et al. 2010a). This is the main goal of two ambitious space mission projects in development:  PLATO (Catala et al. 2010) and TESS (Ricker et al. 2010). Still, another and more straightforward possibility exists. Doppler surveys target bright nearby stars, and they have now  detected enough nearby low-mass planets  to make highly probable that a few of them transit their parent stars. This motivated us to search with {\it Spitzer} for transits of low-mass
Doppler planets having the highest transit probability. In a previous paper (Gillon et al. 2010, hereafer G10), we described the reasons that have led us to conclude that {\it Spitzer} and its Infra-Red Array Camera (IRAC, Fazio et al. 2004) were the best instrumental choice for this transit search, and presented the results of our {\it Spitzer} cycle 5 program targeting HD\,40307\,b (Mayor et al. 2009). The rest of our program consist of a cycle 6 DDT program (ID 60027) of 100 hours that targeted ten other low-mass planets. {\it Spitzer}'s cryogen was depleted at the end of cycle 5, and these observations were thus carried out in non-cryogenic mode (``warm {\it Spitzer}"). 

The recent announcement of 55 Cnc e transits detection by the {\it MOST} satellite (Winn et al. 2011) motivated the publication of this paper. Our initial analysis of the warm {\it Spitzer} data taken in last January concluded a transit detection but also that several sources of instrumental effects needed to be fully characterized before securing the detection. We only recently obtained a satisfactory instrumental model for warm {\it Spitzer} photometry, through a global analysis of calibration data and of all the observations of our cycle 6 program (Gillon et al., in prep.).  
Once applied to our 55 Cnc data, this instrumental model leads not only to the firm detection of the transit of \object{55 Cnc e}, but also to a precise determination of its transit parameters.

Section 2 presents our derivation of the transit ephemeris from the published RVs. In Section~3, we present our data and their analysis that reveals the transiting nature of the planet. We discuss our transit detection and its implications in Section~4.

\section{Transit ephemeris estimation}

We performed a global analysis of all the available RVs for 55\,Cnc to estimate the most reliable transit ephemeris for 55\,Cnc\,e. This analysis was done with the adaptative Markov Chain Monte-Carlo (MCMC) algorithm described in G10. We assumed Keplerian orbits for the five planets orbiting 55\,Cnc, after having checked that  planet-planet interactions had negligible influence on our solutions, using for this purpose the {\it Systemic Console} software package (Meschiari et al. 2009). Our analysis was based on the orbital solution recently presented for 55\,Cnc\,e by Dawson \& Fabrycky (2010).  As shown by these authors, the orbital period value initially reported for this planet, 2.8 days (McArthur et al. 2004; Fischer et al. 2008), was an alias of the true period, 0.74 day.  We verified this result by making two independent MCMC analyses of the RVs, one assuming $P\sim 0.74$ day and the other assuming $P\sim 2.8$ days. 

Using the Bayesian Information Criterion (BIC; e.g. Carlin \& Louis 2008) to estimate the marginal likelihood of both models, and assuming that these models have the same prior probability, we obtained an odds ratio of $\sim10^{16}$ in favor of the P = 0.74 day model, indicating a decisive strength of evidence for this model (Jeffreys 1961). The best-fitting model obtained from this analysis was used to estimate the jitter noise in the RV datasets. 6.0 \ms for Lick, 4.3 \ms for Keck, 5.5 \ms for HET and 15 \ms for ELODIE were added in quadrature to the published error bars to derive the uncertainties on the physical parameters of 55\,Cnc\,e presented in Table~1.

In addition to some basic parameters for the host star,  the origin of the RVs used as input data, and a description of our warm {\it Spitzer} observations, Table~1 provides the most relevant results of our MCMC analysis for 55\,Cnc\,e. The large transit probability, $\sim$29\%, and the very well constrained transit ephemeris (1$\sigma$ error $<$ 1 hour in 2011) of this super-Earth ($M_p = 7.8 \pm 0.6\: M_\oplus$) made it an extremely interesting target for our transit search program.

\begin{table}
\begin{center}
\label{tab:targets}
\begin{tabular}{cc}
\hline\noalign {\smallskip}
Star   & 55\,Cnc  \\ \noalign {\smallskip}
\hline \noalign {\smallskip}
Distance $d$ [parsec]                                 & $12.34 \pm 0.12$$^1$    \\ \noalign {\smallskip}
$V$ magnitude                                           & $5.96 \pm 0.01$$^2$      \\ \noalign {\smallskip}
$K$ magnitude                                           &  $4.02 \pm 0.03$$^3$     \\ \noalign {\smallskip} 
Spectral type                                               & K0V - G8V$^4$        \\ \noalign {\smallskip} 
Effective temperature \teff [K]             & $5234 \pm 30^5$    \\ \noalign {\smallskip} 
Surface gravity \logg                                   & $4.45 \pm 0.08^5$  \\ \noalign {\smallskip} 
Metallicity Fe/H [dex]                                  & $+0.31 \pm 0.04^5$ \\ \noalign {\smallskip} 
Mass $M_\ast$   [$M_\odot$]                      & $0.905 \pm 0.015^6$   \\ \noalign {\smallskip} 
Radius $R_\ast$  [$R_\odot$]            & $0.943 \pm 0.010^6$        \\ \noalign {\smallskip} 
\hline \noalign {\smallskip}
RV data  &      \\ \noalign {\smallskip}
\hline \noalign {\smallskip}
                                                                              & 250 Lick$^4$                         \\ \noalign {\smallskip} 
                                                                              & 70 Keck$^4$                          \\ \noalign {\smallskip} 
                                                                              & 119 HET$^7$                        \\ \noalign {\smallskip} 
                                                                              & 48  ELODIE$^8$                   \\ \noalign {\smallskip}
\hline \noalign {\smallskip}
Planet (MCMC results)  & 55\,Cnc\,e  \\ \noalign {\smallskip}
\hline \noalign {\smallskip}
Minimal Mass $M_p \sin{i}$ [$M_\oplus$]             &  $7.80 \pm 0.56$                     \\ \noalign {\smallskip} 
Expected Radius $R_p$ [$R_{\oplus}$]$^a$                         & 1.3 - 5.7                                  \\ \noalign {\smallskip} 
Expected Area ratio $(R_p/R_\ast)^2$ [ppm]                        &  150 - 3000                              \\ \noalign {\smallskip} 
Equilibrium temperature $T_{eq}$ [K]$^b$             & $1958 \pm 15$                        \\ \noalign  {\smallskip} 
$T_{transit}-2450000$ [HJD]                                 & $5568.011 \pm 0.025$             \\ \noalign {\smallskip}     
$T_{occultation}-2450000$ [HJD]                          & $5568.368 \pm 0.030$             \\ \noalign {\smallskip}     
Orbital period $P$ [d]                                             & $0.7365437 \pm 0.0000052$   \\ \noalign {\smallskip} 
Central transit duration $W_{b=0}$ [min]                & $98 \pm 2$                              \\ \noalign {\smallskip} 
RV semi-amplitude $K$ [\ms]                                 & $5.93 \pm 0.42$                       \\ \noalign {\smallskip} 
Semi-major axis $a$ [AU]                                       &  $0.01544 \pm 0.00009$         \\ \noalign {\smallskip} 
Eccentricity $e$                                                      &  $0.061_{-0.043}^{+0.065}$    \\ \noalign {\smallskip}
Argument of periastron $\omega $ [deg]                & $202_{-70}^{+88}$                \\ \noalign {\smallskip} 
$Prior$ transit probability [\%]                                 & $28.9 \pm 1.5$                        \\ \noalign {\smallskip} 
$Prior$ occultation probability [\%]                          &  $29.3 \pm 1.8$                      \\ \noalign {\smallskip} 
\hline \noalign {\smallskip}
Warm {\it Spitzer} data  &      \\ \noalign {\smallskip}
\hline \noalign {\smallskip}  
Channel  [$\mu$m]                                               & 4.5                                       \\ \noalign {\smallskip} 
AOR$^c$                                                              & 39524608                             \\ \noalign {\smallskip} 
Exposure time [s]                                                  & 0.01                                     \\ \noalign {\smallskip} 
$N_{BCD}$$^d$                                                     & 5240                                     \\ \noalign {\smallskip} 
Duration  [hr]                                                         & 5                                           \\ \noalign {\smallskip}        
 \hline \noalign {\smallskip}    
                                               
\end{tabular}
\end{center}
\caption{Basic data for the star 55 Cnc, relevant results of our MCMC analysis of the RVs, and description of the data (RVs + warm {\it Spitzer} observations) used in this work.
$^1$Van Leeuwen (2007). 
$^2$Turon et al. (1993).
$^3$Skrutskie et al. (2006).
$^4$Fischer et al. (2008).
$^5$Valenti \& Fischer (2005). 
$^6$von Braun et al. (2011). 
$^7$Mac Arthur et al. (2004).
$^8$Naef et al. (2004). 
$^a$Assuming $M_p \sin{i} = M_p$. The minimum and maximum values correspond, respectively, to a pure iron and a pure hydrogen planet (Seager et al. 2007).
$^b$Assuming a null albedo and a heat distribution factor $f'$ = 1/4 (Seager 2010).
$^c$AOR = Astronomical Observation Request = {\it Spitzer} observing sequence.
$^d$BCD = Basic Calibrated Data = block of 64 subarray exposures.
}
\end{table}

\section{Data analysis}

\subsection{Data description}
55\,Cnc was observed by {\it Spitzer} on 6 January 2011 from 9h41 to 14h39 UT. The data consist of 5240 sets of 64 individual subarray images obtained by the IRAC detector at 4.5 $\mu$m with an integration time of 0.01s, and calibrated by the {\it Spitzer} pipeline version S18.18.0. They are available on the  {\it Spitzer} Heritage Archive database\footnote{http://sha.ipac.caltech.edu/applications/Spitzer/SHA} under the form of 5240 Basic Calibrated Data (BCD) files. We first converted fluxes from the {\it Spitzer} units of specific intensity (MJy/sr) to photon counts, then performed aperture photometry on each subarray image with the {\tt IRAF/DAOPHOT}\footnote{IRAF is distributed by the National Optical Astronomy Observatory, which is operated by the Association of Universities for Research in Astronomy, Inc., under cooperative agreement with the National Science Foundation.} software (Stetson, 1987). We tested different aperture radii and background annuli, the best result being obtained with an aperture radius of 3 pixels and a background annulus extending from 11 to 15.5 pixels from the PSF center. The center of the PSF was measured by fitting a Gaussian profile to each image. We discarded the first ten minutes of data to allow the detector to
stabilize. The $x$-$y$ distribution of the measurements was then looked at, and we discarded the few measurements having a very different position than the bulk of the data. For each block of 64 subarray images, we then discarded measurements with discrepant values of flux, background, $x$ and $y$ positions using a $\sigma$ median clipping (5$\sigma$ for the flux and 10$\sigma$ for the other parameters), and the resulting values were averaged, the photometric error being taken as the error on the average flux measurement. At this stage, a 50$\sigma$ slipping median clipping was used on the resulting light curve to discard totally discrepant fluxes.  

Figure 1 shows the resulting raw light curve, and the time-series for the background and the $x$ and $y$ positions. As can be seen in Fig.~1 and Fig.~2, the measured background showed an unusual evolution during the run. It remained stable during $\sim$3.5 hrs,  then it increased abruptly of a few \%, and finally its scatter increased largely. Such a behavior is most probably of instrumental origin. We included this instrumental effect in our data modeling (see below).

\begin{figure}
\label{fig:1}
\centering                     
\includegraphics[width=9cm]{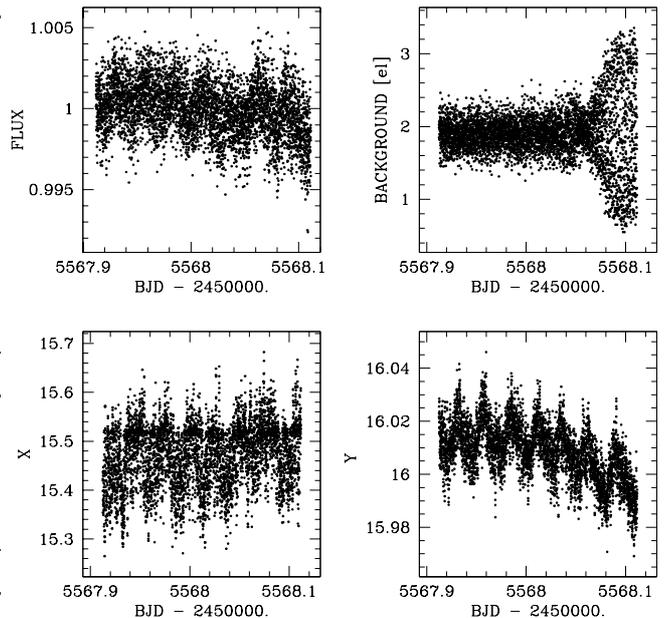}
\caption{{\it Top left}: raw light curve obtained for 55\,Cnc. {\it Top right}: background time-series for this run. {\it Bottom}: time-series for the $x$ ({\it left}) and $y$ ({\it right}) positions of the stellar center.}
\end{figure}

\begin{figure}
\label{fig:2}
\centering                     
\includegraphics[width=9cm]{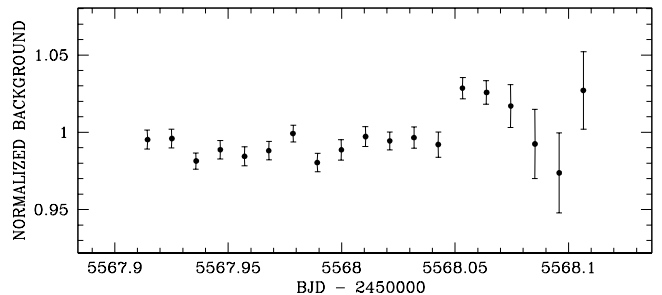}
\caption{Background time-series for the 55\,Cnc data, after normalization and binning per 15 minutes intervals.}
\end{figure}

\begin{figure}
\label{fig:3}
\centering                     
\includegraphics[width=9cm]{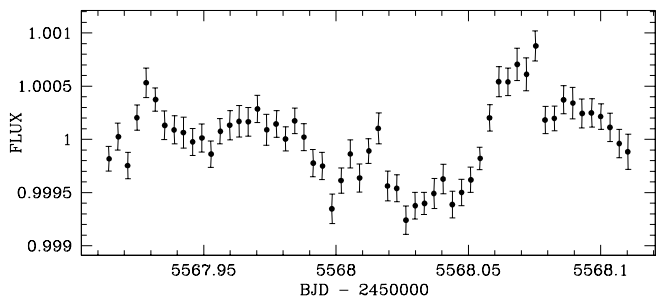}
\caption{55\,Cnc light curve corrected for the ``pixel-phase" effect using a 2$^{nd}$ order position polynomium and binned to intervals of 5 min. }
\end{figure}

\subsection{Modeling the systematics}

The IRAC 3.6 and 4.5 $\mu$m detectors are composed of InSb arrays that show a strong intrapixel quantum efficiency (QE) variability, the QE being maximal in the middle of the pixel and decreasing towards the edges. The full-width at half maximum (FWHM) of the point-spread function (PSF) is $\sim$1.7 pixels. This undersampling of the PSF combined with the QE intrapixel variability leads to a strong dependance of the measured stellar flux on the exact location of the PSF center in a pixel. As {\it Spitzer}'s pointing wobbles with an amplitude of $\sim$0.1 pixel and a period of $\sim$1h, this leads to a severe systematic effect in the photometric time-series acquired at 3.6 and 4.5 $\mu$m, known as the ``pixel-phase" effect. This effect was already present in the cryogenic part of the {\it Spitzer} mission and is very well-documented (e.g., Knutson et al. 2008 and references therein). It is the main limit to the photometric precision of warm {\it Spitzer} (e.g., Ballard et al. 2010). From a comprehensive analysis, the {\it Spitzer} engineering team identified recently the cause of the {\it Spitzer} pointing wobble as the thermal cycling of a heater used to keep a battery within its nominal temperature range\footnote{http://ssc.spitzer.caltech.edu/warmmission/news/ 21oct2010memo.pdf}. After extensive testing and review, it was decided to reduce by a factor of two the thermal amplitude of the cycling while increasing its frequency to make it differ more from the typical frequency of planetary transits and occultations. Our data were obtained after this heater change. The correlation between the measured fluxes and the stellar image position is clearly noticeable in the raw light curve, the resulting periodic pattern having a typical period $\sim$ 35 min, corresponding to a cycle of the heater after the engineering change. We modeled this ``pixel-phase'' effect with the following 2$^{nd}$-order $x$ and $y$ position polynomial: 
\begin{eqnarray}
A(dx,dy) & =  & a_1+a_2dx +a_3dx^2+a_4dy+a_5dy^2 \nonumber\\
              &       & +a_6dxdy \textrm{,}
\end{eqnarray}
where $dx$ and $dy$ are the distance of the PSF center to the center of the pixel. This model for the ``pixel-phase" effect is quite classical in the exoplanet literature (e.g., Knutson et al. 2008, D\'esert et al. 2009). Correcting the light curve with the best-fit ``pixel-phase'' model lead to the light curve visible in Fig.~3. It shows a drop of brightness with an amplitude compatible with a transit of 55\,Cnc\,e. It also shows some other low-amplitude flux modulations that are caused by other warm {\it Spitzer} systematic effects (see below).

One could argue that the transit-like pattern could be caused by the imperfect correction of the ``pixel-phase" effect by the function shown in Eq.~1. This is very unlikely, as the duration of the transit-like structure does not correspond to the one of the wobbles of the stellar position on the chip. To discard firmly this possibility, we tried  3$^{rd}$ and 4$^{th}$-order version of Eq.~1 that led to very similar light curves. We also corrected the ``pixel-phase" effect by a totally different method that relies only on the data themselves and not on any numerical function. We divided the pixel area sampled by the PSF center into $33\times33$ small boxes. If at least 5 subarray measurements felt into a given box, and if these measurements sampled at least 0.14 days (70\% of the duration of the run) the corresponding measurements were divided by their mean value. If these two conditions were not met for a given box, its measurements were discarded. The reduction procedure was then identical to the one described above. The light curve obtained after this correction by an ``intrapixel flatfield" was totally similar (pattern, scatter) to the one visible in Fig.~3. To assess the dependancy of the observed transit-like structure on the details of the reduction procedure several independent reductions of the data were performed by four of us (M. G., B.-O. D., D. D., P. C.), all using different reduction and detrending procedures.  We also tested performing photometry on the 5240 images resulting from the  averaging of the 64 subarray images of each BCD file, using a median filter to reject outlying pixels. Finally, we inspected the light curves obtained without background subtraction. In all cases, the obtained light curves were very similar to the one shown in Fig.~3, confirming the independence of the obtained photometry on the details of the reduction procedure. 

At this stage, we performed a thorough MCMC analysis of our photometry to deduce the transit detection significance, using as input data the raw light curve obtained with an aperture of 3 pixels. Our model assumed a mass of $0.905 \pm 0.015 M_\odot$ for 55\,Cnc (von Braun et al. 2011), and a circular orbit with $P=0.7365437$ days for 55\,Cnc\,e (Sect.~2). We used the model of Mandel \& Agol (2002) for the transit, in addition to the following model for the photometric variations of instrumental and stellar origin:
\begin{eqnarray}
A(dx,dy,dt)  & =  & a_1+a_2dx +a_3dx^2+a_4dy+a_5dy^2 \nonumber\\
                   &    &  + a_6dxdy + a_7dt   \nonumber\\
                   &    &  +  a_8\sin \bigg( \frac{dt - a_{9}}{a_{10}} \bigg) \nonumber\\
                   &    &  + a_{11} \log{dt} + a_{12} \log{dt}^2 \textrm{,}
\end{eqnarray} where $dt$ is the time elapsed since 2455568.05 BJD, i.e. the time at which the background increases sharply (Fig.~1 \& 2). The $a_{11}$ and $a_{12}$ terms were only applied for $dt > 0$. The six first terms of this equation correspond to the ``pixel-phase'' model (Eq.~1). The purpose of the linear term in $dt$ is to model a possible smooth variation of the stellar brightness. The other terms result from our extensive analysis of our entire set of warm {\it Spitzer} data (Gillon et al., in prep.) and of available calibration data that lead us to conclude to a low-amplitude periodic variability of the effective gain of the detector, its typical period lying between 30 and 60 minutes and its amplitude being in average of a few dozens of ppm. Considering the challenging photometric precision required by our program, it is very important to take it into account, justifying the sine term in Eq.~2. We are currently working with the {\it Spitzer} engineering team to find the origin of this periodic variation. We also notice that a ``background explosion'' such as the one affecting the last part of our data is correlated to a sharp increase of the effective gain of the detector that is very well modeled by the last two terms of Eq.~2. The MCMC uses the whole dataset to simultaneously fit for the transit model and the baseline function presented in Eq.~2.

\subsection{MCMC analysis and model comparison}
The following parameters were jump parameters\footnote{Jump parameters are the parameters that are randomly perturbed at each step of the MCMC.} in our analysis:  the planet/star area ratio $(R_p /R_s )^2$, the transit width (from first to last contact) $W$,  the impact parameter $b = a \cos{i}/R_\ast$, and the time of minimum light $T_0$. We assumed a uniform prior distribution for these jump parameters, but we imposed a Gaussian prior for the stellar radius $R_\ast$ based on $R_\ast = 0.943 \pm 0.010 R_\odot$ (Table 1). We assumed a quadratic limb-darkening law with coefficients $u_1=0.0706$ and $u_2=0.1471$. These values were drawn from the theoretical tables of Claret \& Bloemen (2011) for the IRAC 4.5 $\mu$m bandpass and for \teff = 5250 K, \logg =4.5 and [Fe/H]=+0.3. The 12 coefficients of the baseline models (Eq.~2) were determined by least-square minimization at each steps of the Markov chains (see G10 and references therein for details). The correlated noise present in the LC was taken into account as described by G10, i.e., a scaling factor $\beta_r$ was determined from the standard deviation of the binned and unbinned residuals of a preliminary MCMC analysis, and it was applied to the error bars. Several binning intervals ranging from 10 to 90 minutes were tried in preliminary short Markov Chains, and the maximal value for $\beta_r$, 1.35, was used in our analysis.

We performed two new MCMC analyses, one with a transit of 55\,Cnc\,e, and one without. Figure 4 shows the resulting best-fit transit model and its residuals. The odds ratio (Eq.~2 alone) $vs$ (Eq.~2 + transit) is $\sim 10^8$ in favor of the transit model. The transit of 55\,Cnc\,e is thus firmly detected. The period of the sinusoid ($a_{10}$) derived from the MCMC is 51 minutes, well decoupled from the transit duration (96 minutes) and significantly longer than the pixel-phase timescale (35 minutes). Its amplitude is 115$\pm$27 ppm. We show in Fig.~5 the different contributions of the spatially- and time-dependent terms of Eq.~2. This shows that the time-dependent terms are well decoupled from the transit pattern. Table~2 presents the resulting transit and physical parameters and 1$\sigma$ error limits derived for 55\,Cnc\,e. 

\begin{table}[h]
\begin{center}
\begin{tabular}{cc}
\hline \noalign {\smallskip}
$(R_p/R_\ast)^2 $ [ppm]                                       &  $410 \pm 63$                                        \\ \noalign {\smallskip} 
$b=a\cos{i}/R_\ast$ [$R_*$]                                 &  $0.16_{-0.10}^{+0.13}$                            \\ \noalign {\smallskip} 
Transit width  $W$ [d]                                          &   $0.0665_{-0.0019}^{+0.0011}$                \\ \noalign {\smallskip} 
$T_0-2450000$ [BJD]                                          &  $5568.0265_{-0.0010}^{+0.0015}$          \\ \noalign {\smallskip}
$R_p/R_\ast$                                                       & $0.0202_{-0.0016}^{+0.0015}$                 \\ \noalign {\smallskip} 
$a / R_\ast$                                                          & $3.517_{-0.040}^{+0.041}$                       \\ \noalign {\smallskip} 
Inclination $i $ [deg]                                             & $87.3_{-2.1}^{+1.7}$                                 \\ \noalign {\smallskip} 
Radius  $ R_p $ [$R_{\oplus}$]                             & $2.08_{-0.17}^{+0.16}$                           \\ \noalign {\smallskip} 
Mass  $ M_p $ [$M_{\oplus}$]                       & $7.81_{-0.53}^{+0.58}$                           \\ \noalign {\smallskip} 
Mean density $ \rho_p $ [g\,cm$^{-3}$]                   & $4.78_{-1.20}^{+1.31}$                            \\ \noalign {\smallskip} 
\hline \noalign {\smallskip}
\end{tabular}
\caption{Median and 1$\sigma$ limits of the posterior distributions derived for 55\,Cnc\,e from our  MCMC analysis
of our warm {\it Spitzer} photometry. The mass and mean density are derived from the parameters in Table~1.}
\end{center}
\end{table}

We also conducted a residual permutation bootstrap analysis, known as the prayer bead method (Gillon et al. 2006)
to obtain an additional estimation of the residual correlated noise. We used for this purpose the lightcurve corrected from the systematic effects described in Eq.~2. The resulting parameters are in excellent agreement with the ones derived from the MCMC analysis (Table 2), while their error bars are significantly smaller. This result indicates that the error budget is
dominated by the uncertainties on the parameters of the complex baseline model, and not by the residual correlated noise.

To test the robustness of our transit detection and of the resulting transit parameters, we performed $\sim$ 10 additional MCMC analyses as described above, each of them assuming a different set of time-dependent terms presented in Eq.~2. We used a binned lightcurve (per 30s) for the purpose of this comparison to speed up the analysis. Table~3 presents the baseline model, derived depth, BIC (Bayesian Information Criterion) and Bayes factor obtained for 4 of those MCMC analyses.

\begin{table*}[h]
\begin{center}
\begin{tabular}{lccccc}
\hline \noalign {\smallskip}
                                                  & $p$                  & $p + t$                       & $p + s + t$                  & $p + j + t$         & $p + s + j + t$   \\ \noalign {\smallskip}
\hline \noalign {\smallskip}
$(R_p/R_\ast)^2 $ [ppm]      & $590 \pm 72$ &  $665 \pm 70$         &  $683 \pm 87$          &  $428 \pm 62$ & $410 \pm 63$          \\ \noalign {\smallskip} 
BIC                                           &          804          &  758                          &     758                            &       746                &         732          \\ \noalign {\smallskip} 
Bayes factor                           &                           &     $9.8\times10^9$  &   $9.8\times10^{9}$  &   $3.9\times10^{12}$ &  $4.3\times10^{15}$     \\ \noalign {\smallskip} 
\hline \noalign {\smallskip}
\end{tabular}
\caption{Transit depth, Bayesian Information Criterion (BIC) and Bayes factor from the MCMC obtained for 5 different model baselines. Model terms are described as follows : $p$ is the pixel phase correction ($a_2$, $a_3$, $a_4$, $a_5$ and $a_6$ in Eq.~2), $t$ is the time-dependent linear trend ($a_7$), $s$ is the sinusoidal ($a_8$, $a_9$ and $a_{10}$) and $j$ is the jump model ($a_{11}$ and $a_{12}$). The Bayes factor given in the table is relative to the $p$ model. Our adopted model described in Eq.~2 is the rightmost one.}
\end{center}
\end{table*}

While none of these models revealed to be better than our nominal model for representing our warm {\it Spitzer} data (Bayes factor between $10^3$ and $10^{15}$), each of these models lead to a decisive detection of the transit of 55\,Cnc\,e (Bayes factor between $10^{10}$ and $10^{50}$). For all these alternatives models, the deduced values for the transit parameters agreed well with the ones deduced in our nominal analysis, except for the transit depth when the jump is not included. Nevertheless, our Bayesian model comparison makes these alternative models $>4\times10^5$ times less probable than our nominal model. Table 3 also illustrates how the jump and the sinusoidal variation terms improve the baseline model. We thus not only conclude to our firm detection of a transit of 55\,Cnc\,e, but also to the robustness of the deduced results shown in Table~2.

\begin{figure}
\label{fig:4}
\centering                     
\includegraphics[width=9cm]{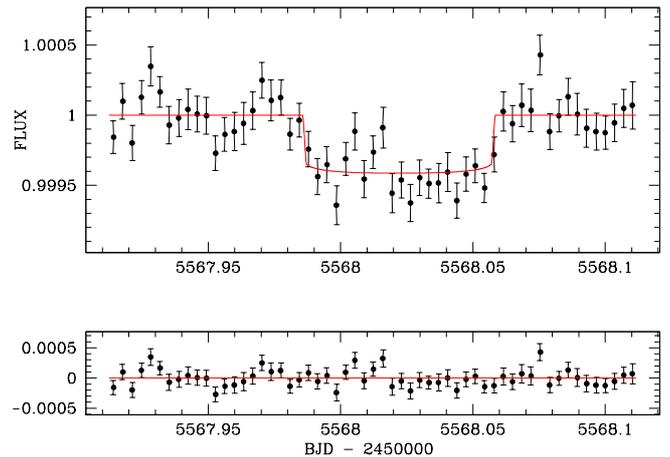}
\caption{$Top$: 55\,Cnc  light curve divided by the best-fit baseline model, binned to intervals of 5 min, with the best-fit transit model overimposed. $Bottom$: residuals of the fit binned to intervals of 5 min. }
\end{figure}

\begin{figure}
\label{fig:5}
\centering                     
\includegraphics[width=9cm]{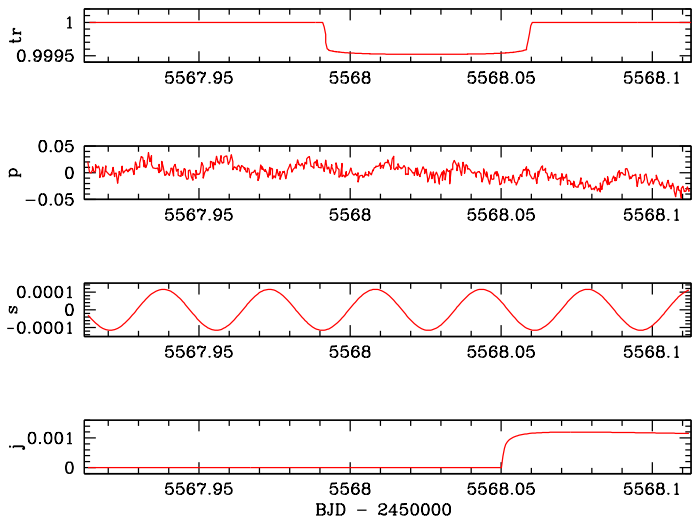}
\caption{Spatially- and time-dependent terms of the model function used in Eq.~2. Model terms are described as follows : $tr$ is the transit model, $p$ is the pixel phase correction ($a_2$, $a_3$, $a_4$, $a_5$ and $a_6$ in Eq.~2), $s$ is the sinusoidal ($a_8$, $a_9$ and $a_{10}$) and $j$ is the jump model ($a_{11}$ and $a_{12}$). The linear trend $a_7$ is not represented.}
\end{figure}

\section{Discussion and conclusions}

\subsection{Photometric precision of warm {\it Spitzer}} 
Because we have to take into account three different instrumental effects in addition to 
a possible smooth variation of the stellar flux, the complexity of our photometric baseline 
model is large (12 free parameters, Eq.~2). This illustrates well the challenge of 
ultra-precise time-series IR photometry, especially with a detector that
is no longer actively cooled. By modeling this baseline in addition to the 
transit in our MCMC analysis, we naturally take into account its uncertainties
and their impact on the deduced transit parameters. Despite the complexity of 
the baseline model, we reach a very good precision on these transit parameters. 
This is due not only to the extensive characterization of the warm {\it Spitzer} 
instrumental effects performed by the exoplanet community and the {\it Spitzer} Science Center, 
but also by the extremely high-cadence made possible by the IRAC subarray mode. 
Indeed, we have here more than 5,000 photometric measurements to constrain thoroughly 
the 12+4 parameters of our global model. We show here that warm {\it Spitzer} can
not only detect an eclipse of a few hundreds of ppm (Ballard et al. 2010), it can also measure its depth 
with a  precision of $\sim60$ ppm, leading to the conclusion that this space telescope 
has still an important role to play for the detection and characterization of 
transiting planets.

\subsection{Planetary radius}

Our MCMC results (see Table~2) yield a planetary radius of $2.08_{-0.16}^{+0.17}\: R_{\oplus}$ as measured in IRAC 4.5$\mu$m channel. The error bars are determined from the posterior distribution function produced by the MCMC and includes the error on the stellar radius. The current planetary radius uncertainty is dominated by the error on the transit depth. 

On its side, the radius of the star itself is now extremely well constrained, thanks to recent interferometric observations of 55 Cancri performed by van Braun et al. (2011) using the CHARA array. The resulting updated stellar radius value (Table~1) now yields a negligible contribution from the stellar radius to the planetary size uncertainty.

In the preprint version of this paper, we reported a planetary radius 30\% larger than the one initially obtained by Winn et al. (2011) in the visible with MOST. After we submitted our paper, a new version of the Winn et al. (2011) analysis was made available that yields good agreement with our results (at the 1$\sigma$ level).

\subsection{Composition of 55\,Cnc\,e}

We used the internal structure
model described in Valencia et al. (2010) suitable for rocky and gaseous
planets. We considered four different rocky compositions that span
the possible range in radius. The upper bound for the radius is set
by the lightest rocky composition, which is one where there is no
iron. A planet with a radius larger than this upper limit necessarily
has volatiles. The lower bound for the radius is set by a pure iron
composition. Both extreme compositions are unlikely to exist given
that 1) iron, magnesium and silicate have similar condensation temperatures,
with the latter two making up most of the mantle of the Earth (i.e.
Mg$_{0.9}$Fe$_{0.1}$O+ SiO$_{2}$) and 2) a pure iron composition
is unlikely even with maximal collisional stripping (Marcus et al. 2010). The other two rocky compositions
are an Earth-like one (33\% iron core, 67\% silicate mantle with 0.1
of iron by mol) and a ``super-Mercury'' (63\% iron core, 37\% silicate
mantle no iron). We also consider volatile compositions in which we
added different amounts of H$_{2}$O or hydrogen and helium (H-He)
at an equilibrium temperature of $\sim$2000 K above an earth-like
nucleus. 

The data obtained in this study for the radius ($2.08^{+0.16}_{-0.17}\: R_\oplus$), and mass ($7.81_{-0.53}^{+0.58}\: M_\oplus$) place the composition of 55\,Cnc\,e intersecting the threshold line between planets that necessarily require volatiles
(above the ``no-iron" line), and the ones that may be rocky (below the ``no-iron" line). However, most of the combinations of mass and radius lie above the upper limit of a
rocky planet, requiring that 55\,Cnc\,e have volatiles in its composition. 
We find that an envelope of a few parts in 10,000 of H-He or of order of $\sim$10\% water above
an Earth-like core can fit the data well. In Fig.~6 we show the mass-radius relationships for the
different compositions considered and the different known transiting
super-Earths. Based on the same arguments
proposed by Valencia et al. (2010) for CoRoT-7\,b, the timescale for evaporation
of a H-He envelope would be too short ($\sim$ a few million years) for it to
be considered as a plausible composition, whereas the timescale for
water evaporation is of the same order of magnitude than that of the
age of the system ($\sim$ a few billion years). Thus, according to
the \textit{Spitzer} data analysed in this study, we favor a composition of
an envelope of supercritical water above a solid, perhaps earth-like,
nucleus. The exact amount of volatiles will depend on the composition
of the solid nucleus, with a reasonable estimate around $\sim$15\%. However, a pure rocky composition cannot be ruled out, in which case the planet would be depleted in iron with respect to Earth. 

Similarly, the data for 55\,Cnc\,e reported by Winn et al. 2011 also lies at the threshold of these two types of planets, albeit with a denser composition.   

Figure 7 shows the density as a function of mass of
several transiting super-Earths.  While CoRoT-7\,b and Kepler-10\,b have practically the same composition, 55\,Cnc\,e, with its similar effective temperature and mass, has a much lighter composition. It lies between the high-density ``super-Mercuries" and the volatile-rich planets Kepler-11\,b and GJ\,1214\,b. Within a small range of masses, 4-9 $M_\oplus$, the known transiting super-Earths span a relatively large variety in compositions.

\begin{figure}
\centering
\includegraphics[width=9cm]{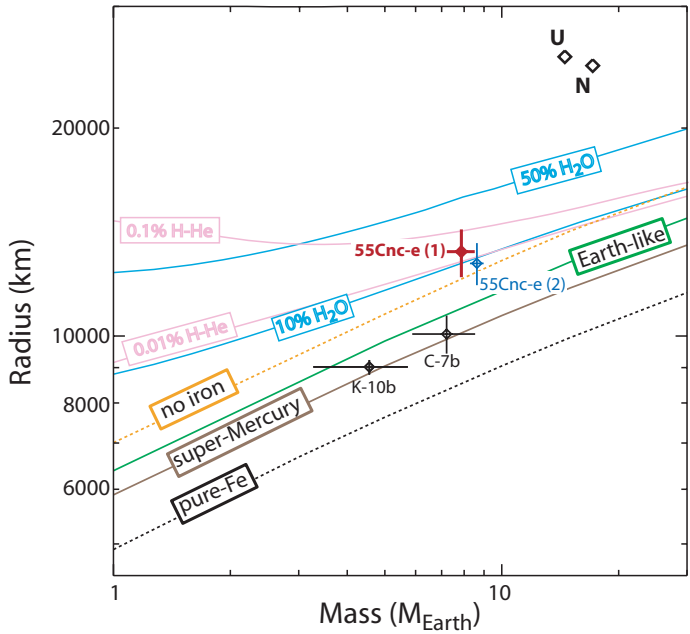}\label{fig:6}
\caption{Mass-Radius relationship for 55\,Cnc\,e. We show four different rocky
compositions: no iron, Earth-like (33\% iron core, 67\% silicate mantle
with 0.1 of iron by mol), super-Mercury (63\% iron core, 37\% silicate
mantle no iron), and a pure iron planet. We consider two types of
volatiles compositions: 0.1-0.01\% of H-He (pink), and 10-50\% of water
(blue) above an Earth-like nucleus. We show our data for 55\,Cnc\,e (red
cross with label (1)), that reported by Winn et al. (2011) (blue and label (2)), and the data for the
known transiting hot super-Earths, Kepler-10\,b (K-10b -- data from
Batalha et al. 2011), and CoRoT-7\,b (C-7b data from Bruntt et al. 2010,
Hatzes et al. 2011). Uranus and Neptune are shown for reference. }
\end{figure}

\begin{figure}
\centering
\includegraphics[width=9cm]{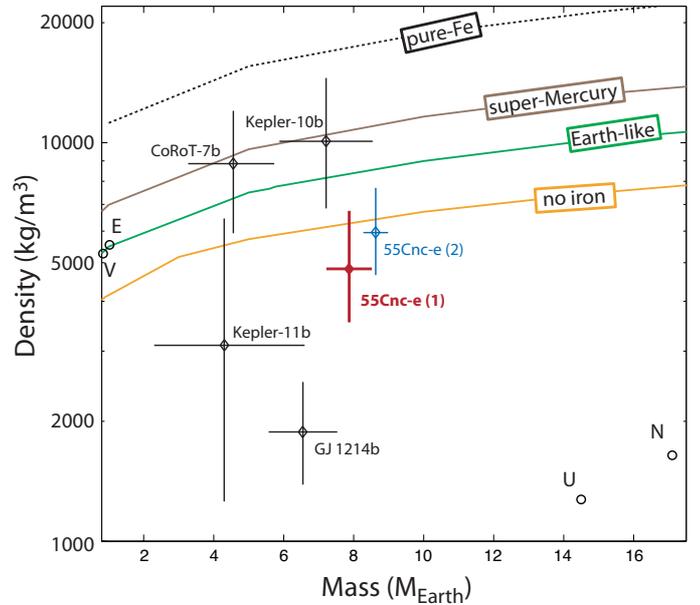}\label{fig:7}

\caption{Density vs mass of transiting super-Earths. The data for the known
transiting super-Earths and mini-Neptunes are shown, as well as the
relationships for the four rocky compositions considered in this study.
Earth, Venus, Uranus and Neptune are shown for reference. The error bars for the density are calculated as $\rho_{min}=4\pi/3\,(M-\sigma M)/{(R+\sigma R)^3}$ and $\rho_{max}=4\pi/3\,(M+\sigma M)/{(R-\sigma R)^3}$. }
\end{figure}

\begin{acknowledgements} 
We are grateful to K. von Braun and S. Kane for insightful discussions on the stellar radius of 55\,Cnc. We thank the anonymous referee for an helpful review that improved the paper.
This work is based in part on observations made with the {\it Spitzer Space Telescope}, which is operated by the Jet Propulsion Laboratory, California Institute of Technology under a contract with NASA. Support for this work was provided by NASA. We thank the {\it Spitzer} Science Center staff for efficient scheduling of our observations. M. Gillon is FNRS Research Associate. This publication makes use of data products from the Two Micron All Sky Survey, which is a joint project of the University of Massachusetts and the Infrared Processing and Analysis Center/California Institute of Technology, funded by the National Aeronautics and Space Administration and the National Science Foundation.
\end{acknowledgements} 

\bibliographystyle{aa}

\begin{thebibliography}{}

   \bibitem[2010]{Bakos} Bakos G. \'A., Torres G., P\'al A., et al. 2010, ApJ, 710, 1724

   \bibitem[2010]{Ballard} Ballard S., Charbonneau D., Deming D., et al., 2010, PASP, 122, 1341

   \bibitem[2008]{Barge} Barge P., Baglin A., Auvergne M., et al., 2008, A\&A, 482, L17

   \bibitem[2011]{Batalha} Batalha N. M., Borucki W. J., Bryson S. T., et al., 2011, ApJ, 729, 27
  
   \bibitem[2010]{Bean} Bean J. L., Miller-Rici Kempton E., Homeier D., 2010, Nature, 468, 669
   
   \bibitem[2010]{Borucki1} Borucki W. J., Koch D. G., Basri G., et al., 2010a, Science, 327, 977
  
   \bibitem[2010]{Borucki2} Borucki W. J., Koch D. G., Brown T. M., et al., 2010b, ApJ, 713, L126

   \bibitem[2011]{Borucki3} Borucki W. J., Koch D. G., Basri G., et al. 2011, ApJ (submitted), arXiv:1102.0541
   
   \bibitem[2010]{Bruntt} Bruntt H., Deleuil M., Fridlund M., et al., 2010, A\&A, 519, A51

   \bibitem[2008]{carlin} Carlin B.~P., Louis T.~A., 2008, Bayesian Methods for Data Analysis, Third Edition (Chapman \& Hall/CRC)   
  
   \bibitem[2011]{Carter} Carter J. A., Winn J. N., Holman M. J., et al., 2011, ApJ (accepted), arXiv:1012.0376
  
   \bibitem[2010]{Catala} Catala C., Arentoft T., Fridlund M., et al., 2010, Proceeding of the meeting Pathways Towards Habitable Planets. Eds. Coud\'e du Foresto V. , Gelino D. M., Ribas I. San Franciso. ASP, 260. 

   \bibitem[2009]{Charbonneau} Charbonneau D., Berta Z.~K., Irwin J., et al., 2009, Nature, 462, 891
   
   \bibitem[2011]{Claret} Claret  A., Bloemen S., 2011, A\&A, 529, A75

   \bibitem[2010]{Dawson10} Dawson R. I., Fabrycky D. C., 2010, ApJ,  722, 937

   \bibitem[2009]{Desert} D\'esert, J.~M., Lecavelier des Etangs A., H\'ebrard G., et al., 2009, ApJ, 699, 478

   \bibitem[2011]{Desert2} D\'esert, J.~M., Bean J., Miller-Ricci Kempton E., et al., 2011, ApJ, 731, L40

   \bibitem[2004]{Fazio} Fazio G. G., Hora J. L., Allen L. E.,  et al., 2004, ApJS, 154, 10

   \bibitem[2007]{Fischer07} Fischer D. A., Vogt S. S., Marcy G. W., et al., 2007, ApJ, 669, 1336

   \bibitem[2008]{Fischer08} Fischer D. A., Marcy G. W., Butler R. P., et al., 2008, ApJ, 675, 790

   \bibitem[1996]{Flower} Flower P.~J., 1996, ApJ, 469, 355

   \bibitem[2006]{Gillon3} Gillon M., Pont F., Moutou C., et al., 2006, \aap, 459, 249
  
   \bibitem[2007]{Gillon} Gillon M., Pont F., Demory B.-O., et al, 2007, A\&A, 472, L13

   \bibitem[2010]{Gillon2} Gillon M., Deming D., Demory B.-O., et al., 2010, A\&A, 518, A25

   \bibitem[2010]{Hartman} Hartman J. D., Bakos G. \'A., Kipping D. M., et al., 2010, ApJ, 728, 138
  
   \bibitem[2010]{Hatzes} Hatzes A.P., Dvorak R., Wuchterl G., et al., 2010, A\&A, 520, A93

   \bibitem[1961]{Jeffreys} Jeffreys H., 1961, The Theory of Probability, Oxford University Press
 
   \bibitem[2008]{Knutson} Knutson H.\~A., Charbonneau D., Allen L.\~A., et al., 2008, ApJ, 673, 526

   \bibitem[2009]{Leger} L\'eger A., Rouan D., Schneider J., et al., 2009, A\&A, 506, L287

   \bibitem[2011]{Lissauer} Lissauer J. J., Fabrycky D. C., Ford E. B., et al., 2011, Nature, 470, 53

   \bibitem[2009]{Lovis} Lovis C., Mayor M., Bouchy F., et al., 2009, Transiting Planets, Proceedings of the International Astronomical Union, IAU Symposium 253, 502

   \bibitem[2002]{Mandel} Mandel K., Agol E., 2002, ApJ, 580, 171 

   \bibitem[2010]{Marcus} Marcus, R.~A., Sasselov, D., Hernquist, L., \& Stewart, S.~T.\ 2010, \apjl, 712, L73 

   \bibitem[2009]{Mayor1} Mayor M., Udry S., Lovis C., et al., 2009, A\&A, 493, 639
 
   \bibitem[2004]{McArthur} McArthur B. E., Endl M., Cochran W. D., et al., 2004, ApJ, 614, L81

   \bibitem[2009]{Meschiari} Meschiari S., Wolf A.~S., Rivera E., et al., 2009, PASP, 121, 1016
 
   \bibitem[2004]{Naef} Naef D., Mayor M., Beuzit J.~L., et al., 2004, A\&A, 414, 351

   \bibitem[2006]{Rafikov} Rafikov R. R., 2006, ApJ, 648, 666
   
   \bibitem[2010]{Ricker} Ricker G. R., Latham D. W., Vanderspek R. K., et al., 2010, AAS Meeting 215, Bulletin of the American Astronomical Society, 42, 459

   \bibitem[2010]{Rogers} Rogers L. A., Seager S., 2010, ApJ, 716, 1208

   \bibitem[2007]{Seager07} Seager S., Kuchner M., Hier-Majumder C.~A., Militzer B., 2007, ApJ, 669, 1279

   \bibitem[2010]{Seager10} Seager S., 2010, Exoplanet Atmospheres, Princeton University Press
   
   \bibitem[2011]{Shabram} Shabram M., Fortney J. J., Greene T. P., Freedman R. S., 2011, ApJ, 727, 65
   
   \bibitem[2006]{Skrutskie} Skrutskie M. F., Cutri R. M., Stiening R., et al., 2006, AJ, 131, 1163

   \bibitem[1987]{Stetson} Stetson P.~B., 1987, PASP, 99, 111

   \bibitem[2010]{Sumi} Sumi T., Bennett D. P., Bond I. A., et al., 2010, ApJ, 710, 1641

   \bibitem[2007]{Torres} Torres G., 2007, ApJ, 671, L65

   \bibitem[1993]{Turon} Turon C., Creze M., Egret D., et al., 1993, Bull. Inf. Centre Donnees Stellaires, 43, 5
   
   \bibitem[2009]{vanBelle} van Belle G.~T., von Braun K., 2009, ApJ, 694, 1085

   \bibitem[2007]{VanLeeuwen} Van Leeuwen F., 2007, A\&A, 474, 653
       
   \bibitem[2010]{Valencia} Valencia, D., Ikoma, M., Guillot, T., \& Nettelmann, N. 2010, A\&A, 516, A20
   
   \bibitem[2005]{Valenti} Valenti J. A., Fischer D. A., ApJS, 159, 141
   
   \bibitem[2011]{vonBraun11} von Braun, K., et al.\ 2011, ApJ Letters (submitted), arXiv:1106.1152 

   \bibitem[2011]{Winn} Winn J.~N., Matthews J.~M., Dawson R., et al., 2011, ApJ Letters (submitted), arXiv:1104.5230
 
\end{thebibliography}

\end{document}